\documentclass[letterpaper]{article} 
\usepackage{aaai23}  
\usepackage{times}  
\usepackage{helvet}  
\usepackage{courier}  
\usepackage[hyphens]{url}  
\usepackage{graphicx} 
\usepackage{natbib}  
\usepackage{caption} 
\usepackage{algorithm}
\usepackage{algorithmic}

\usepackage{subfiles} 
\usepackage{pifont}
\newcommand{\cmark}{\ding{51}}%
\newcommand{\xmark}{\ding{55}}%
\usepackage{makecell}
\usepackage{multirow}
\usepackage{graphics}
\usepackage{adjustbox}
\usepackage{booktabs}
\usepackage{amsmath}
\usepackage{newfloat}
\usepackage{listings}
\DeclareCaptionStyle{ruled}{labelfont=normalfont,labelsep=colon,strut=off} 
\floatstyle{ruled}
\newfloat{listing}{tb}{lst}{}
\floatname{listing}{Listing}
\title{A Domain-Knowledge-Inspired Music Embedding Space and a Novel Attention Mechanism for Symbolic Music Modeling}

\author {
    Zixun Guo\textsuperscript{\rm 1}\thanks{This paper is accepted at AAAI 2023.}
    \thanks{This work was supported by Singapore MOE Tier 2 Grant No. MOE2018-T2-2-161.},
    Jaeyong Kang\textsuperscript{\rm 1},
    Dorien Herremans\textsuperscript{\rm 1}
}
\affiliations {
    \textsuperscript{\rm 1} Information Systems Technology and Design, Singapore University of Technology and Design, Singapore\\
    \{nicolas\_guo, jaeyong\_kang, dorien\_herremans\}@sutd.edu.sg
}

\usepackage{bibentry}
\begin{document}
\maketitle

\begin{abstract}
Following the success of the transformer architecture in the natural language domain, transformer-like architectures have been widely applied to the domain of symbolic music recently. Symbolic music and text, however, are two different modalities. Symbolic music contains multiple attributes, both absolute attributes (e.g., pitch) and relative attributes (e.g., pitch interval). These relative attributes shape human perception of musical motifs. These important relative attributes, however, are mostly ignored in existing symbolic music modeling methods with the main reason being the lack of a musically-meaningful embedding space where both the absolute and relative embeddings of the symbolic music tokens can be efficiently represented. In this paper, we propose the Fundamental Music Embedding (FME) for symbolic music based on a bias-adjusted sinusoidal encoding within which both the absolute and the relative attributes can be embedded and the fundamental musical properties (e.g., translational invariance) are explicitly preserved. Taking advantage of the proposed FME, we further propose a novel attention mechanism based on the relative index, pitch and onset embeddings (RIPO attention) such that the musical domain knowledge can be fully utilized for symbolic music modeling. Experiment results show that our proposed model: RIPO transformer which utilizes FME and RIPO attention outperforms the state-of-the-art transformers (i.e., music transformer, linear transformer) in a melody completion task. Moreover, using the RIPO transformer in a downstream music generation task, we notice that the notorious degeneration phenomenon no longer exists and the music generated by the RIPO transformer outperforms the music generated by state-of-the-art transformer models in both subjective and objective evaluations. The code of the proposed method is available online\footnote[1]{\url{github.com/guozixunnicolas/FundamentalMusicEmbedding}}
\end{abstract}

\section{Introduction}\label{sec:introduction}


The transformer architecture~\cite{transformer} has recently achieved remarkable successes in the natural language domain. Since symbolic music and text are both sequential, the transformer architecture, with its ability to handle global structures, has gradually replaced the recurrent neural network (RNN) and its variants (e.g., GRU, LSTM) in sequential modeling~\cite{music_transformer,cp_transformer,remi,herremans2017functional, briot2020deep}. Inspired by the language pretraining models such as Word2Vec and BERT~\cite{word2vec, BERT}, several attempts have been made to discover the semantic meaning of the symbolic music tokens (SMTs)~\cite{musicbert,musebert,pirhdy}. Symbolic music and text, however, are different modalities. Simply adapting the transformer-based language modeling or pretraining techniques to the symbolic music domain will lead to the loss of rich inductive biases of music (i.e., music domain information) and does not guarantee that the trained embedding space accurately models the fundamental music properties. We use the first few bars of the jazz standard: ``Giant Steps'' as an example to better illustrate the difference between symbolic music and text in Figure~\ref{fig:difference}. In the Figure, we include the sheet music and the most basic form of the event-based representation of the sheet music which contains a series of pitch, duration and onset tokens. The index (i.e., ordering) of the tokens is also included. For simplicity, the performance level attributes (e.g., velocity, timbre) are omitted.

\begin{figure}
    \centering
    \includegraphics[width = \linewidth]{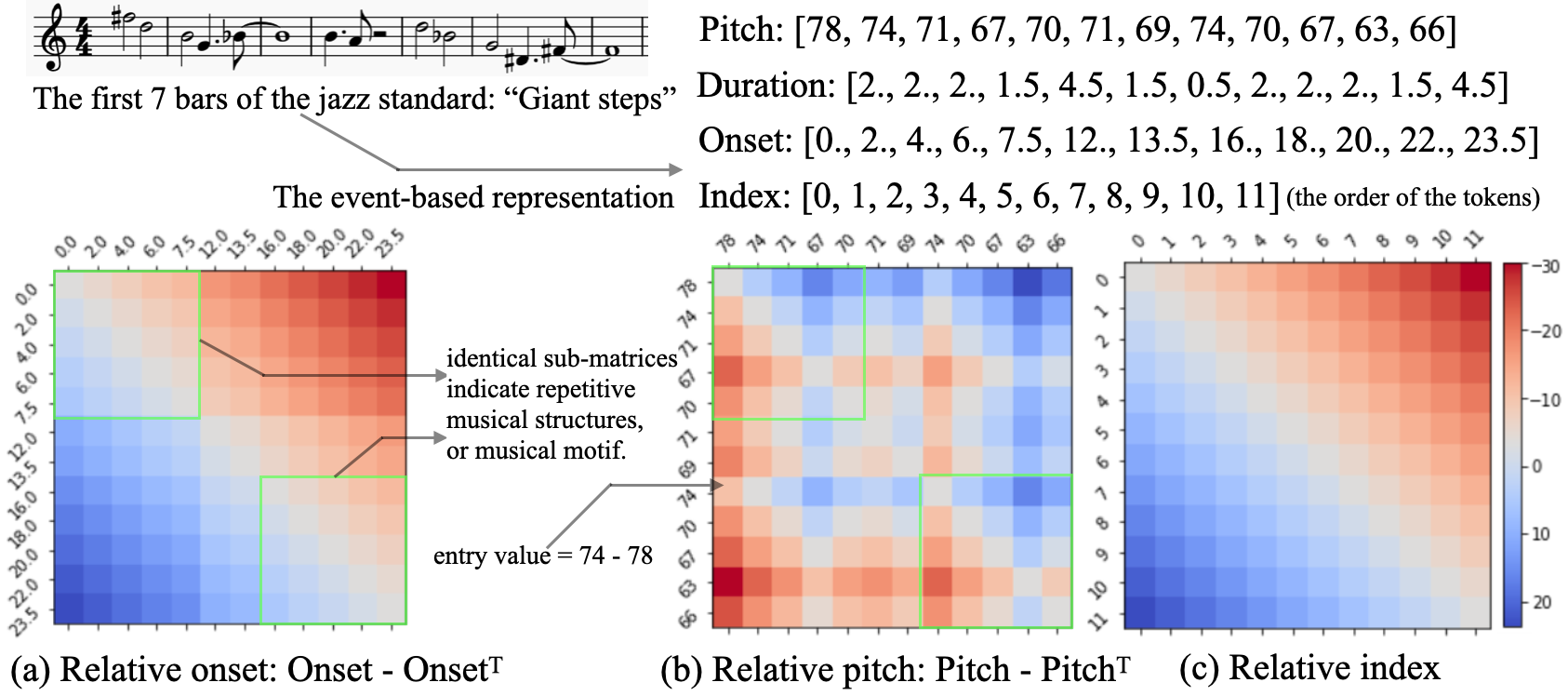}
    \caption{Sheet music and the event representation of the jazz standard ``Giant Steps''. Rich musical domain information (i.e., motif) is reflected in the relative attributes.}
    \label{fig:difference}
\end{figure}


We start by listing three main differences in modeling language versus symbolic music. Firstly, since the transformer is not position-aware, a common practice to indicate the positions of the texts is to use an index-based sinusoidal positional encoding~\cite{transformer}. The index-based positional encoding, however, is not able to fully capture the positioning of the symbolic music since the positioning of SMTs by nature is onset-and-beat-based. For example in Figure~\ref{fig:difference}, describing the position of the first D note as ``the note at the 3rd beat of the 1st bar'' is more precise than ``the 2nd note in the note sequence''. Secondly, a translational invariance property should exist in a musical embedding space which cannot be guaranteed using common language pretraining techniques~\cite{word2vec, BERT} or end-to-end trainable embeddings. For example, the distance (interval) between any pitch pairs a major second apart (e.g., C4 and D4; F4 and G4) in the embedding space should be unique and identical. Thirdly, it is noteworthy to mention the importance of the relative onset and pitch attributes which shape the human perception of musical motifs. For instance, the core mechanism of the motif-discovering algorithm COSIATEC~\cite{cosiatec} is based on the relative pitch and onset. A cognitive study~\cite{cog_sci1} also came to a similar conclusion that humans perceive music ``incrementally'' which confirms the importance of the relative attributes of music. For example, the relative attributes of the example music (i.e., relative pitch, onset and index) are 
obtained by calculating the self-distance matrices and are shown in Figure~\ref{fig:difference} a, b and c respectively. The iconic motif is hence revealed in the identical sub-matrices (labelled in green) of the relative onset and pitch attributes. The relative index attribute indicates the relative positions of the SMTs yet provides little domain information of the music as compared to the relative pitch and onset. Using common language modeling techniques where only the absolute attributes (i.e., pitch, duration and onset) are included as input, however, would not reveal this important domain information. For transformers to include this important musical inductive bias, these relative attributes along with the original attributes need to be embedded. Yet, designing or training such musically-meaningful embedding space where the absolute embeddings (e.g., embeddings for pitch) and relative embeddings (e.g., embeddings for intervals) exist concurrently is nontrivial. Hence, existing transformer architectures designed for symbolic music either discard this important inductive bias~\cite{cp_transformer,remi} or apply an end-to-end trainable relative embedding solely on the relative index (Figure~\ref{fig:difference} c) during relative attention calculation~\cite{music_transformer}. \citet{music_transformer}, however, did mention the use of the relative pitch and onset embeddings but declared the method ``not scalable beyond the J.S. Bach chorale dataset''. Even though additional trainable embeddings of relative pitch and onset can be included and trained end-to-end, the relationship between the relative and the original embedding space remains unexplainable.

In this paper, we introduce the Fundamental Music Embedding (FME), a domain-knowledge-inspired embedding space within which the embeddings of musical tokens (e.g., pitch) and their relative embeddings (e.g., interval) co-exist and the fundamental music properties are preserved explicitly. Inspired by the distance-aware and easy-to-transpose property of the position encoding (PE) function~\cite{transformer}, we utilize the deterministic sinusoidal encoding (SE) function of the PE as the backbone of the FME and propose to add trainable biases to the SE such that the FME spaces of different types of SMTs (e.g., pitch, duration) and the relative embeddings (e.g., interval and time shift) can be easily distinguished. We further take advantage of the FME and propose a novel attention mechanism that captures the relative index, pitch and onset (RIPO attention) for transformer-based symbolic music modeling.

In our extensive experiments, our proposed model: RIPO transformer which utilizes FME and RIPO attention outperforms the state-of-the-art (SOTA) transformers (i.e., music transformer and linear transformer) equipped with various embedding methods in a melody completion task in terms of cross-entropy loss. In a downstream music generation task, our proposed RIPO transformer successfully tackles the notorious degeneration problem. The music generated by the RIPO transformer also outperforms those generated by the baseline models in both subjective and objective evaluations. 



\section{Related Work}\label{sec:related_work}


\subsection{Semantically-meaningful music embedding spaces}

The spiral array model~\cite{chew2000towards} is a tonality model that models the relationship between notes, chords and key centres in a 3-dimensional space. Moving to the deep learning era where symbolic music is often modelled by RNN architectures~\cite{eckblues,this_time,CMHRNN, tonnetz}, the main approach to embed music tokens is to use one-hot or multi-hot encoding. In one-hot embedding spaces, the embedded vectors are orthogonal to each other. Yet, each one-hot embedded vector is equidistant from all other vectors in the embedding space thus causing ambiguity. Representing the relative embeddings of musical attributes (e.g., pitch interval) in one-hot embedding spaces is non-trivial. To represent the relative embeddings, one has to utilize rotational matrices to distinguish different one-hot vector pairs or use convolutional neural networks to capture the intervals between pitches~\cite{tonnetz}. The size of one-hot or multi-hot embeddings, however, will grow with an increasing input vocabulary size and the sparse representation will negatively affect the computational efficiency.

With the advent of language pretraining techniques~\cite{word2vec}, similar attempts have been made to discover the semantic meaning of music~\cite{cotext_to_concept, chordripple,melody2vec}. Variational autoencoders (VAE) have also been utilized to extract latent representations of symbolic music~\cite{pianotree,musicvae}. For example, the shared embedding space between pitch and duration of the PianotreeVAE~\cite{pianotree} has been analyzed and translational invariance of note duration, as well as regular pitch patterns, are observed which conveys the fundamental music properties. These methods mostly operate on a higher level (i.e., melody, chord). In this paper, however, we aim to discover the semantic representation of symbolic music on a more fundamental level (i.e., note-level).

In recent years, transformer-like architectures have become a game-changer in sequential modeling. The music transformer~\cite{music_transformer} is the very first work that utilizes relative attention~\cite{shaw-rpe} for music generation. It utilizes one-hot embeddings to encode the SMTs. \citet{music_transformer} emphasize the importance of the relative attributes of symbolic music and a novel ``skewing'' operation is proposed to efficiently calculate the relative index embedding (Figure~\ref{fig:difference}c) for symbolic music sequences. In this paper, we aim to further improve the relative attention calculation mechanism by taking into account the relative pitch and onset embedding (Figure~\ref{fig:difference}a, b) which provides more musical domain information and can be directly inferred from our proposed FME without involving additional trainable embedding spaces. The music transformer has been extended to a conditional generation model~\cite{KPOP_transformer}. In~\cite{remi}, a novel REMI representation of symbolic music has been proposed and the original transformer architecture has been replaced with transformer-XL~\cite{transformerxl} to deal with the extra-long music sequences. The REMI representation has further been improved~\cite{cp_transformer} and a linear transformer architecture~\cite{linear_transformer} is utilized to increase computation efficiency. \citet{remi,cp_transformer} utilize end-to-end trainable embeddings for symbolic music tokens and the relative attributes of symbolic music are no longer utilized. By studying the trained embedding spaces, we notice that the translational-invariance property of symbolic music cannot be observed. Several methods have been proposed to use BERT-like architectures to achieve contextualized music representation learning~\cite{pirhdy,musicbert,musebert}. Even though promising results have been achieved in downstream MIR tasks, interpreting the raw embedding spaces and the contextualized embeddings is challenging and the relative embeddings of symbolic music cannot be represented explicitly. Whereas in our proposed FME, the fundamental music properties can be interpreted and the relative embeddings can be directly inferred.


\subsection{Sinusoidal encoding}
We review the sinusoidal encoding (SE), also known as the positional encoding proposed by \citet{transformer} on which our Fundamental Musical Embedding (FME) is based. The SE is able to transform an index-based sequence: $I: \{0, ..., n-1\}$ into a sequence of vectors that conveys the index order in a $d$-dimensional space $E_I \in \mathcal{R}^{n\times d}$. As a result, the transformer becomes position-aware. The SE has the following properties: 1. The L2 distance between token pairs in the embedding space is translational-invariant; 2. Any token in the embedded space can be converted to any other token using a deterministic linear operation; 3. The embedding space is continuous such that SE can be interpolated and extrapolated to other non-integer or negative inputs. We find the first property useful in designing a musically-meaningful embedding space. Since we aim to represent both the absolute and relative embedding in the FME, the second property provides a promising blueprint. Moreover, the third property facilitates the transition from index-based position encoding for text to the proposed onset-and-beat-based positional encoding for symbolic music. 

\section{Fundamental Music Embedding}\label{sec:fme}
A basic symbolic music sequence  of length $n$ is defined as a series of pitch ($p$), duration ($d$) and onset ($o$) triplets: $\{(p_1, d_1, o_1),...(p_n,d_n, o_n) \}$. This sequence can be regrouped into three subsequences for pitch, duration and onset respectively: $P:\{p_1, ..., p_n\}$, $D:\{d_1, ..., d_n\}$, $O:\{o_1, ..., o_n\}$. See Figure~\ref{fig:difference} for a detailed example. More generally, we define such subsequences as sequences of fundamental music tokens (FMTs) $F:\{f_1, ..., f_n\}$. The relative attribute of FMT is defined as dFMT: $\Delta F:\{min(F)-max(F), ..., max(F)-min(F)\}$ which represents all the possible relative differences in $F$. We define $P, D, O, F \in \mathcal{R}^{n\times 1}$ as the vector representation of their respective token type. The embedding function for FMT: $F$ is defined as the Fundamental Music Embedding function (FME) $FME_F:\mathcal{R}^{n\times 1}\rightarrow\mathcal{R}^{n\times d}$. The embedding function for dFMT is defined as the Fundamental Music Shift (FMS). In other words, FME represents the absolute embedding and FMS represents the relative embedding. The FMTs, FMSes and FMEs should observe the following properties in order to preserve the fundamental musical properties:


\begin{enumerate}

    \item Translational invariance: $|f_a-f_b|=|f_c-f_d|	\Rightarrow 
\| FME_F(f_a)-FME_F(f_b)\|_2= \| FME_F(f_c)-FME_F(f_d)\|_2, \forall f_a,f_b,f_c,f_d \in F$ 
    \item Transposability: $FME_F(f_{a+k}) = G(FME_F(f_a), FMS_F(k))$, where $G$ is a linear function and $k \in \Delta F$ represents the transpose value. $FMS_F$ should be directly inferrable from the $FME_F$.  

    \item Separability: in the FME space, different types of embedded FMTs (e.g., embeddings of pitch and duration) and their relative embeddings (e.g., embeddings of interval and time shift) should be well separated.

\end{enumerate}

We use a type of FMT: pitch ($p$) as an example to interpret the properties listed above. Property 1 preserves the interval relationship among any pitch pairs in the embedding space such that any pitch pair with the same interval: $|a-b|$ or $|c-d|$ will have the same L2 distance in the embedding space. In other words, distance in the FME relates to musical intervals. Property 2 indicates that there exists an explicit representation of relative embeddings (i.e., embeddings of pitch intervals) and these relative embeddings can be utilized to transpose one pitch to another in the embedding space. For instance, a relative pitch embedding of an ``ascending major third'' can be utilized to transpose each pitch by a major third in the embedding space. As opposed to FMTs, there also exists non-FMTs (e.g., the `pad' token) which do not contain the aforementioned 3 properties. For these non-FMTs, a normal end-to-end trainable embedding will be applied.


We start by formulating our proposed FME. A $d$-dimensional Fundamental Music Embedding function $FME_F$ for FMT $f \in F$ is defined in Equation~\ref{eq:Ex}. For simplicity, $FME_F$ is abbreviated as $FME$ where $F$ represents the type of FMT (e.g., pitch, duration). It consists of a total of $\frac{d}{2}$ sub-components $P_k(f)$ which are defined in Equation~\ref{eq:Px}. The sub-component $P_k(f)$ consists of a sinusoidal vector:$[sin(w_k f), cos(w_kf)]$ where $w_k$ is an exponentially decreasing function controlled by a base value: $B$ defined in Equation~\ref{eq:wx} and a trainable bias vector: $[b_{sin}k, b_{cos}k]$. To obtain multiple types of FME for various types of FMTs (e.g., pitch, duration), a different base value $B$ can be chosen for $w_k$. Due to the nature of the Fourier transform and the exponential function, the sinusoidal vectors for different types of FME in Equation~\ref{eq:Px} will always be orthogonal to each other, thus fulfilling Property 3 (Separability) in Section~\ref{sec:fme}. The key difference between the positional encoding~\cite{transformer} and the proposed FME is the usage of trainable bias terms. The purpose of using trainable bias terms is also to ensure Property 3 (Separability) since different types of FMTs will be separable in the embedding space by the biases. A detailed proof for Property 1 (Translational Invariance) in Section~\ref{sec:fme} is provided in the Appendix.

\begin{equation}\label{eq:wx}
    w_k = B^{-\frac{2k}{d}}
\end{equation}
\begin{equation}\label{eq:Px}
    P_k(f) = [sin(w_k f)+b_{sin}k, cos(w_kf)+b_{cos}k]
\end{equation}
\begin{equation} \label{eq:Ex}
    FME(f) = [P_0(f), ..., P_k(f), ... P_{\frac{d}{2}-1}(f)]
\end{equation}

From the FME space, the relative embeddings FMS for the relative attribute (i.e., transpose value) dFMT $\Delta f \in \Delta F$ can be explicitly represented. Similarly, we define a sub-component of FMS: $A_k(\Delta f)$ in Equation~\ref{eq:Sx}. Note that the same exponentially decreasing function $w_k$ is utilized but the trainable bias vector is not included. The FMS is formed by concatenating its sub-components $A_k(\Delta f)$ and is defined in Equation~\ref{eq:Tx}. The FMS can be utilized to transpose embedded music tokens in the FME space thus fulfilling Property 2 (Transposability) in Section~\ref{sec:fme}. A detailed proof of this is provided in the Appendix.




\begin{equation} \label{eq:Sx}
    A_k(\Delta f) = [sin(w_k\Delta f), cos(w_k\Delta f)]
\end{equation}
\begin{equation} \label{eq:Tx}
    FMS(\Delta f) = [A_0(\Delta f), ..., A_k(\Delta f), ... A_{\frac{d}{2}-1}(\Delta f)]
\end{equation}



\section{RIPO attention: attention mechanism that uses relative index, pitch and onset embeddings}

\begin{figure*}[h]
    \centering
    \includegraphics[width = 0.7\linewidth]{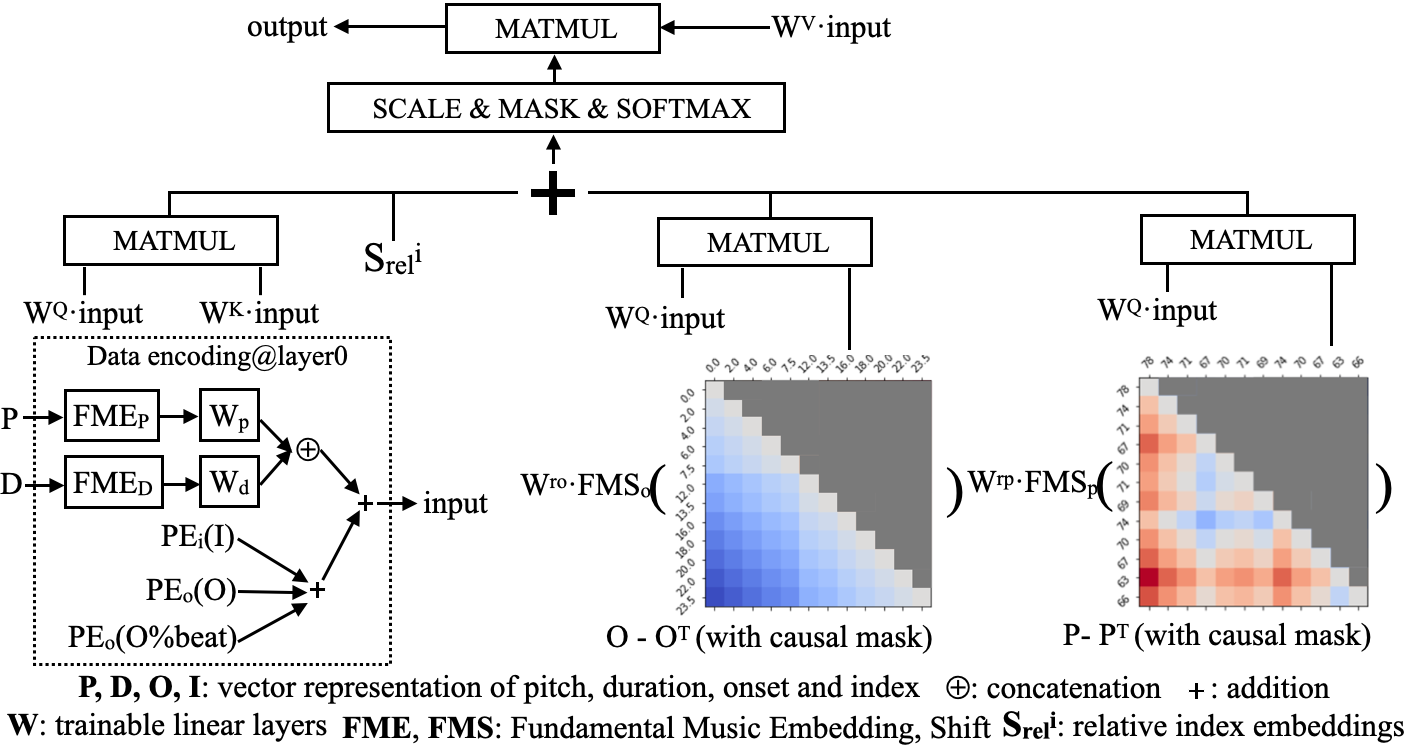}
    \caption{RIPO attention layer.}
    \label{fig:rpd_trans}
\end{figure*}

In this section, we formulate a novel attention mechanism calibrated to symbolic music modeling which incorporates explicit relative index, pitch and onset in the attention calculation (RIPO attention). The current proposed architecture supports monophonic music and we aim to extend this to a polyphonic setting in the future. The architecture of the RIPO attention layer is shown in Figure~\ref{fig:rpd_trans}. Compared to a multi-head attention layer or relative global attention layer, the RIPO attention layer is improved in 2 ways: 1. To better reflect the nature of the onset-and-beat based symbolic music data, two additional positional encodings (PE) based on onset and beat attribute have been added in addition to the original index-based positional encoding. 2. To better reflect the implicit structure of symbolic music, relative onset and pitch embeddings are integrated during attention calculation in addition to the relative index embedding used in the music transformer~\cite{music_transformer}. 

Following the notation from Section~\ref{sec:fme}, we represent a basic symbolic music sequence with length $n$ as: $P,D,O$ where $P, D, O \in \mathcal{R}^{n\times 1}$ represents pitch, duration and onset respectively. The index (i.e., ordering) of the sequence is defined as $I = \{0, ..., n-1\}$. The calibrated PE for symbolic music is illustrated in Equation~\ref{eq:pe} where $PE_i$ is the original index-based positional encoding of the transformer and $i$ indicates the PE is applied to the indices. We propose to add two additional $PE_o$s based on the absolute onset $O$ and the musical beat $O\%beat$. Here, $beat$ represents the number of beats per bar, $o$ indicates the PE is applied for onsets. The modulo operation \% acts like a `metronome'. 



\begin{equation}\label{eq:pe}
    PE = PE_i(I) + PE_o(O) + PE_o(O \% beat)
\end{equation}

The input to the RIPO transformer is defined in Equation~\ref{eq:input} where $FME_P$ and $FME_D$ are the proposed FME for pitch and duration respectively. Since we only focus on monophonic music modeling for now, the onset sequence $O$ can be ignored as the onsets for monophonic music can be obtained by calculating the cumulative sum of the duration input. We use $\boldsymbol{W_p}$ and $\boldsymbol{W_d}$ to represent the two trainable linear layers and $\oplus$ is a vector concatenation operation. The $input$ is further projected to the query ($Q$), key($K$) and value($V$) vectors using three trainable linear layers: $\boldsymbol{W_Q}$, $\boldsymbol{W_K}$, $\boldsymbol{W_V}$ in Equation~\ref{eq:qkv}. 

\begin{equation}\label{eq:input}
    input = \boldsymbol{W_p}FME_P(P)\oplus \boldsymbol{W_d}FME_D(D)+PE
\end{equation}
\begin{equation}\label{eq:qkv}
    Q,K,V = \boldsymbol{W_Q}input, \boldsymbol{W_K}input, \boldsymbol{W_V}input
\end{equation}

The RIPO attention calculation is defined in Equation~\ref{eq:Srelp}-\ref{eq:attn}. The relative attention logits: $S_{rel}^P$ and $S_{rel}^O$ for pitch and onset are obtained in Equation~\ref{eq:Srelp}-\ref{eq:Srelo}. $FMS_P$ and $FMS_O$ are two different types of relative embeddings for relative pitch and onset which can be directly inferred from their respective FME space. $\boldsymbol{W^{rp}}$ and $\boldsymbol{W^{ro}}$ are 2 trainable linear layers and are applied to the relative pitch and onset embeddings respectively. The output of the attention layer is defined in Equation~\ref{eq:attn} where we inherit the skewing operation to calculate the relative index embedding: $S_{rel}^i$ from~\citet{music_transformer}. $D_h$ represents the number of hidden dimensions per attention head~\cite{transformer}.



\begin{equation}\label{eq:Srelp}
    S_{rel}^P=Q[\boldsymbol{W^{rp}}FMS_P(P-P^T)]^T
\end{equation}
\begin{equation}\label{eq:Srelo}
    S_{rel}^O= Q[\boldsymbol{W^{ro}}FMS_O(O-O^T)]^T
\end{equation}
\begin{equation}\label{eq:attn}
    output = softmax(\frac{QK^T+S_{rel}^i+S_{rel}^P+S_{rel}^O}{\sqrt{D_h}})V
\end{equation}

\section{Experimental setup}\label{sec:experiment_setup}


\subsection{Dataset and pre-processing}
To validate the effectiveness of the proposed RIPO transformer which utilizes FME and RIPO attention for symbolic music modeling, we start from the basics and use monophonic symbolic music datasets for training. In future research, we aim to extend the current architecture into a polyphonic setting. We define the training objective as a melody completion task (i.e., next-note prediction) which can be used as a music generation model subsequently. We have collected 10,199 pieces of music with a time signature of 4/4 from the TheoryTab\footnote{\url{https://www.hooktheory.com/theorytab}} and Wikifonia dataset\footnote{\url{http://www.synthzone.com/files/Wikifonia/}}, and randomly selected 90\% for training and 10\% for testing. All data are transposed into the key of C major or A minor to reduce complexity. The maximum sequence length allowed is 246 and music sequences shorter than this will be padded in the end. We limit the smallest duration quantization unit to be a 16th note (duration value: 0.25) and the largest duration to be a whole note (duration value: 4.0). Hence, a total of 16 unique duration units will be included in the duration dictionary. Duration tokens with values greater than a whole note are represented with a `sustain token' which is a non-FMT token to mimic the behaviour of a musical tie and to reduce the vocabulary size for duration tokens. For instance, a music token [(C4, 5.0)] will be split into two tokens: [(C4, 4.0), (sustain, 1.0)].

\subsection{Baseline models and embeddings} 
We choose two transformer models for symbolic music modeling as the baseline models: 1. music transformer~\cite{music_transformer} which utilizes relative index embeddings~\cite{shaw-rpe} during attention calculation; 2. linear transformer~\cite{linear_transformer} which is utilized in the compound-word transformer~\cite{cp_transformer}. As our dataset is monophonic and less complex, we notice using 2 attention layers results in the lowest training and testing loss for both the baselines and our proposed model and hence keep 2 attention layers for all the models. Moreover, to validate the effectiveness of the proposed FME and FMS, we choose 3 types of embedding as our baselines: 1. OH: one-hot encoding; 2. WE: end-to-end trainable word embedding; 3. W2V: pretrained word embedding using Word2Vec pretraining method~\cite{word2vec}; 4.W2V freeze:  same as W2V but frozen during training. We use a window size of 2 and 4 negative tokens per positive token to pretrain the W2V embedding~\cite{word2vec}. The cross-entropy (CE) losses for the pitch token, duration token, as well as the total losses will be compared. The implementation details (e.g., hyper-parameters) are provided in the Appendix. 


\subsection{Music generation and listening test setup}
We utilize the trained models in a downstream music generation task. We select the first 2 bars of music as the seed inputs from every song in the test set and generate a total of 16 bars of music for each seed input. We observe a serious degeneration phenomenon~\cite{degeneration} using the baseline models. To be more specific, the generated music easily falls into endlessly repeating loops which conforms to the observation from the compound word transformer online code repository. To offset this, we utilize temperature-controlled top-k and top-p sampling during music generation. Besides the subjective evaluation, a listening test which includes 25 participants is conducted to compare the generated results (30 pieces of music) from the RIPO transformer and the best baseline model with the same seed inputs. The participants are required to rate the music snippets on the following aspects on a 5-point Likert scale: 1. the overall enjoyment; 2. the correctness in terms of pitch; 3. the correctness in terms of duration; 4. the interestingness of the generated music. 

\begin{table}
  \caption{Cross entropy (CE) loss comparison by ablating each relative embedding and each positional encoding.}
  
  \centering
\begin{adjustbox}{width=0.8\linewidth,center}\label{tab:ablation}

  \begin{tabular}{ccccccccc}
    \toprule
    No. & $S_{rel}^o$ &$S_{rel}^p$& $S_{rel}^i$ & $PE(O)$ &$PE(O\%beat)$ &$CE_p$ &$CE_d$ &$CE_{sum}$\\   
    \midrule
    1     &\cmark &\cmark &\xmark &\cmark &\cmark & 1.463& 0.935& 2.398  \\
    2     &\cmark &\xmark &\cmark &\cmark &\cmark & 1.451& 0.931 & 2.381  \\ 
    3      &\xmark &\cmark &\cmark &\cmark &\cmark & 1.630& 1.056& 2.686     \\ 
    4     &\xmark &\xmark &\cmark &\cmark &\cmark & 1.626& 1.053& 2.679  \\ 
    \midrule
    5    &\cmark &\cmark &\cmark &\xmark &\cmark &  1.477& 0.936& 2.413 \\
    6     &\cmark &\cmark &\cmark &\cmark &\xmark &  1.466& 0.936& 2.402 \\ 
    7      &\cmark &\cmark &\cmark &\xmark &\xmark &  1.466& 0.933& 2.400    \\ 
    \midrule
    8      &\cmark &\cmark &\cmark &\cmark &\cmark & \textbf{1.440}& \textbf{0.927} & \textbf{2.367}    \\ 
    \bottomrule
  \end{tabular}
\end{adjustbox}
\end{table}

\section{Results}
\subsection{Onset-and-beat-based PE and RIPO attention}


We evaluate the effectiveness of the proposed RIPO attention and the onset-and-beat-based position encoding via an ablation experiment. In Table~\ref{tab:ablation}, we first ablate the relative embeddings in Models No.1-4 and ablate the positional embeddings in Models 5-7 subsequently and the cross entropy losses (CE loss) of these models are reported. Note that the original index-based position encoding $PE_i$ is kept for all models. Table~\ref{tab:ablation} confirms the effectiveness of our approach as the proposed RIPO transformer (Model No. 8) with the most inductive biases (i.e., all the proposed relative embeddings and positional encodings) achieves the lowest CE loss on the test set.

  


\subsection{RIPO transformer+FME versus baseline models}

Table~\ref{tab:comparison} shows that our proposed RIPO transformer which utilizes the proposed FME and the RIPO attention mechanism has outperformed all baseline transformers equipped with different embedding methods in a melody completion task in terms of CE loss. Additionally, we have two novel findings. Firstly, unlike recent transformer architectures~\cite{cp_transformer,remi} where end-to-end trainable embeddings are applied to the musical tokens, we notice that replacing these embeddings with one-hot embedding will result in lower CE losses for both music transformer (MT) and linear transformer (LT). Secondly, MT outperforms LT regardless of embedding methods in general. This may be caused by the additional use of relative index embeddings. 

\begin{figure}
    \centering
    \includegraphics[width = 0.9\linewidth]{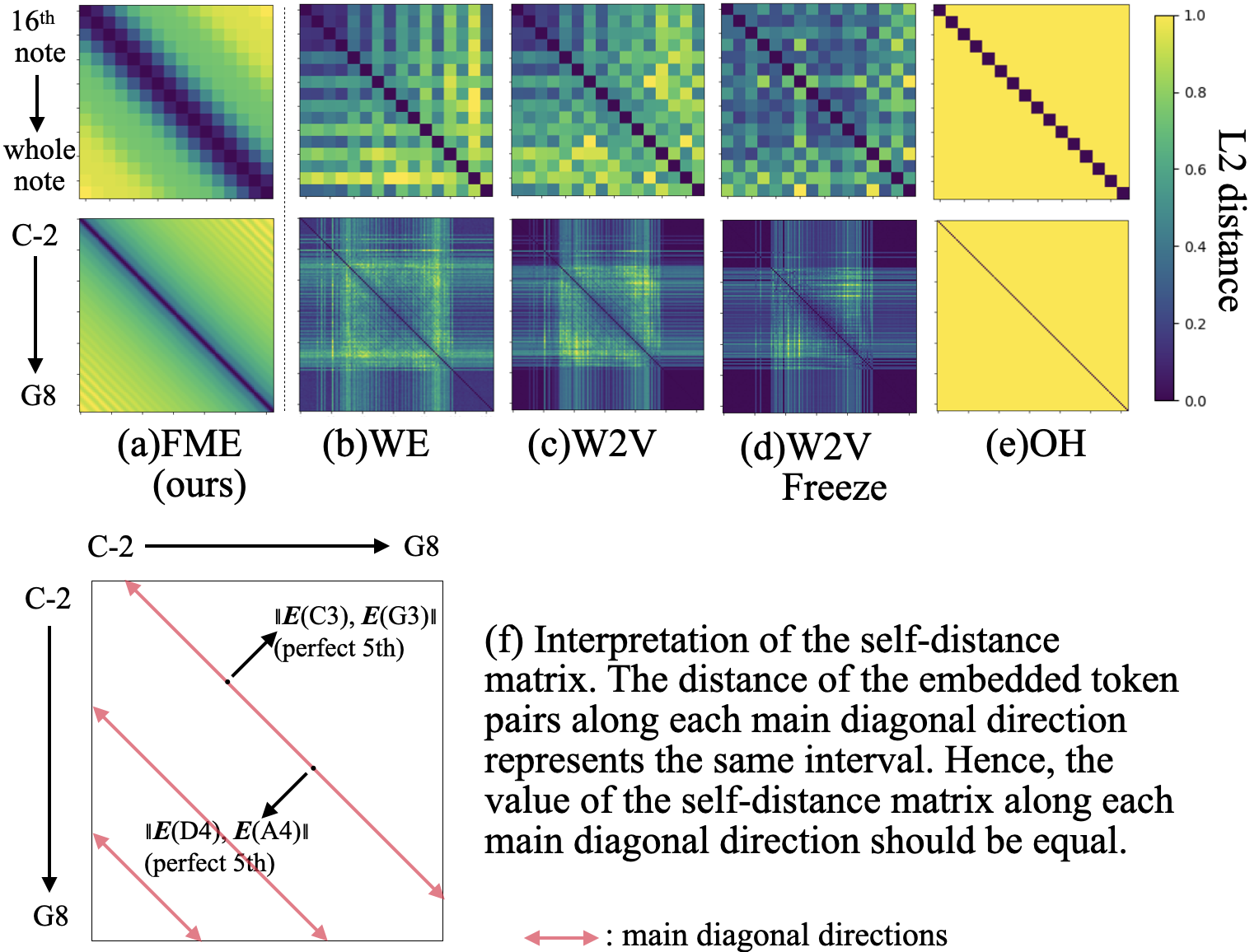}
    \caption{Self-distance matrices of the embedded pitch and duration tokens using different embedding methods.}
    \label{fig:visual}
\end{figure}

To examine whether the translational invariance property exists in the embedding space, the self-distance matrices of the embedded musical tokens using different embedding methods are calculated and plotted in Figure~\ref{fig:visual}. More specifically, a self-distance matrix is obtained by calculating the L2 distance for all embedded token pairs: $\| E(X)-E(X)^T \|_2$ where $X$ are pitch or duration tokens arranged in increasing orders and $E$ is an embedding function. Results in Figure~\ref{fig:visual} show that the translational invariance property only holds for the proposed embedding function: FME and one-hot embedding. More specifically, L2 distances of tokens pairs along the main diagonal direction (red arrows in Figure~\ref{fig:visual} f) remain the same for FME and one-hot encoding but differ for other end-to-end trainable embeddings. This might be one of the reasons that FME and one-hot embeddings outperform other embedding methods in Table~\ref{tab:comparison}. Yet, any two one-hot vectors (except two identical ones) are equidistant in the embedding space. Hence Figure~\ref{fig:visual} e shows a uniform color except for the diagonal. This does not reflect distances between different pitch pairs. For example, the distance between C4 and E4 and the distance between C4 and G5 are considered equal in the one-hot embedding space. 



\begin{table*}[h!]
  \caption{Model comparison and the objective metrics of the generated music. We use MT and LT to refer to music transformer and linear transformer respectively. WE, W2V, and OH stands for word embedding, word2vec, and one-hot encoding respectively. KL, ISR, and AR stands for KL divergence, in-scale ratio, and arpeggio ratio respectively.}
  \label{tab:comparison}
  \centering
\begin{adjustbox}{width=0.95\linewidth,center}
  \begin{tabular}{cllcccccccccccc}
    \toprule
    \multicolumn{8}{c}{model configuration}&\multicolumn{3}{c}{test loss}&\multicolumn{4}{c}{objective evaluation}\\
    \cmidrule(lr){1-8}\cmidrule(lr){9-11}\cmidrule(lr){12-15}\\
    No. &\makecell[l]{ Embed. method}& Model &$S_{rel}^o$ &$S_{rel}^p$& $S_{rel}^i$ & $PE_o$ &$PE_b$ &$CE_p$ &$CE_d$ &$CE_{sum}$&$KL_p$&$KL_d$&$ISR$&$AR$\\   
    \midrule
    1  &WE& MT  &\xmark &\xmark &\cmark &\xmark &\xmark & 1.452& 0.956& 2.408&0.014&0.039&0.973&0.036  \\
    2  &W2V& MT  &\xmark &\xmark &\cmark &\xmark &\xmark & 1.482& 0.979& 2.461&0.016&0.036&0.970&0.039  \\ 
    3  &{W2V freeze}&  MT  &\xmark &\xmark &\cmark &\xmark &\xmark &  1.484& 0.974& 2.458&0.016&0.035&0.966&0.038    \\ 
    4  &OH& MT   &\xmark &\xmark &\cmark &\xmark &\xmark & 1.450& 0.954& 2.405&0.015&0.035&0.969&0.038  \\ 
    \midrule
    5  &WE& LT &\xmark &\xmark &\xmark &\xmark &\xmark & 1.810& 1.133 &2.943&0.026&0.040&0.971&0.032 \\
    6  &W2V& LT  &\xmark &\xmark &\xmark &\xmark &\xmark &  2.014&1.228 &3.242&0.030&0.067&0.975&0.023\\ 
    7  &W2V freeze&  LT  &\xmark &\xmark &\xmark &\xmark &\xmark &3.008&1.663&4.670&0.056&0.098&\textbf{0.996}&0.006  \\ 
    8  &OH&  LT  &\xmark &\xmark &\xmark &\xmark &\xmark &    1.758& 1.122&2.880 &0.041&0.033&0.972&0.031\\ 
    \midrule
    9  &FME (ours)& RIPO (ours)  &\cmark &\cmark &\cmark &\cmark &\cmark &  \textbf{1.440}& \textbf{0.927}& \textbf{2.367 }& \textbf{0.011}& \textbf{0.024}& 0.981& \textbf{0.049} \\ 

    \bottomrule
  \end{tabular}
\end{adjustbox}
\end{table*}

\subsection{Demystifying the `endless loop' problem during music generation}


\begin{table}[h!]
  \caption{Comparison of seq\_rep attributes.}
  \label{tab:seqrep}
  \centering
\begin{adjustbox}{width=0.9\linewidth,center}

  \begin{tabular}{lllcc}
    \toprule
    Model  &   \makecell[l]{ sampling \\method} &\makecell[l]{ sampling \\temperature}& \makecell[l]{ seq\_rep\_4 \\pitch} & \makecell[l]{ seq\_rep\_4 \\duration}  \\   
    \midrule
    MT+WE    &\multirow{5}{*}{top\_p = 0.9}&\multirow{5}{*}{1.0}&0.713&0.809 \\
    MT+W2V   &&&0.668&0.788 \\ 
    MT+W2V$_{freeze}$ &&&0.666&0.782   \\ 
    MT+OH   &&&0.697&0.791 \\ 
    RIPO+FME (ours)   &&&\textbf{0.294}&\textbf{0.535}   \\ 

    \midrule
    MT+WE    &\multirow{5}{*}{top\_k = 5}&\multirow{5}{*}{1.0}&0.567&0.678 \\
    MT+W2V   &&&0.501&0.644 \\ 
    MT+W2V$_{freeze}$ &&&0.512&0.654  \\ 
    MT+OH   &&&0.531&0.666 \\ 
    RIPO+FME (ours)   &&&\textbf{0.310}&\textbf{0.440}   \\ 
    \midrule

    ground truth data  &-&-&0.328&0.536   \\ 
    \bottomrule
  \end{tabular}
\end{adjustbox}
\end{table}

\begin{figure}[h!]
    \centering
    \includegraphics[width = 1.0\linewidth]{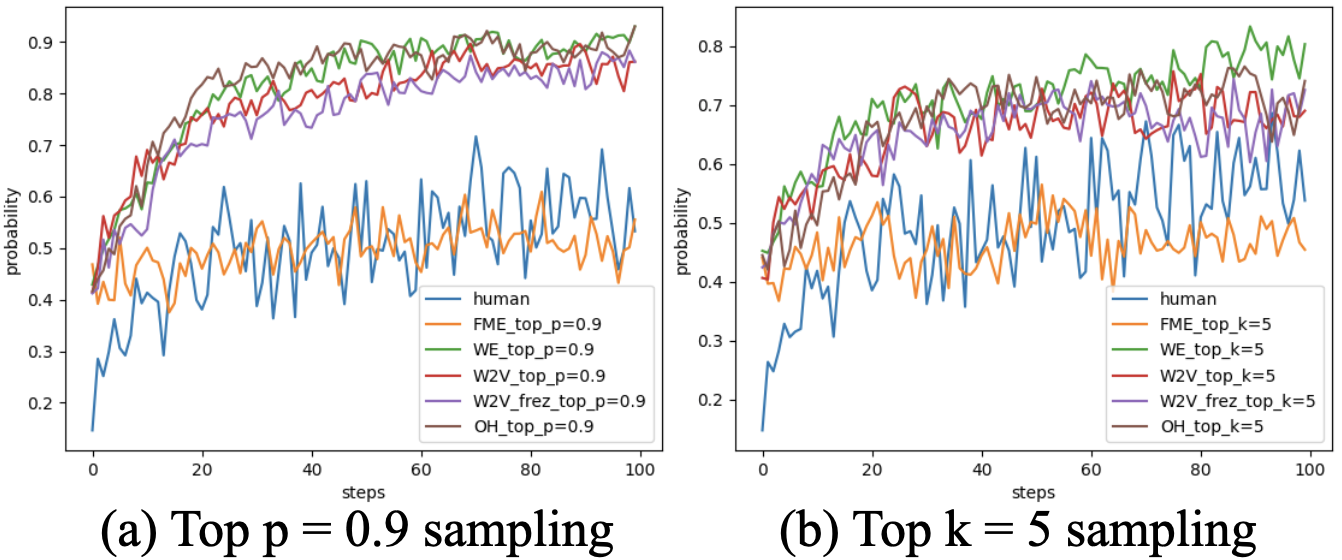}
    \caption{Step-wised probability assigned for different models and human (ground truth) during generation.}
    \label{fig:probility_sampling}
\end{figure}

 In text-generation tasks, texts sampled from trained language models have been observed to fall into endlessly-repeating patterns which is also known as `degeneration'~\cite{degeneration}. This phenomenon has also been observed in music generation tasks~\cite{cp_transformer}. In our experiment, we observe the same degeneration effect in the results generated from the baseline models but not our proposed models. Since the MT consistently outperforms LT in terms of CE loss in Table~\ref{tab:comparison}, we use Models No.1-4 in Table~\ref{tab:comparison} as the baseline models to illustrate the degeneration phenomenon. A $seq-rep-n$ attribute~\cite{seqrep} (ratio of non-unique n-grams to all n-grams) can be calculated to reflect the degeneration level. By setting $n=4$, we calculate the $seq-rep-4$ attribute of the generated music from different models (using top\_k and top\_p sampling with the same sampling temperature of $1.0$) and the ground truth in Table~\ref{tab:seqrep}. In Table~\ref{tab:seqrep}, we observe unnaturally high $seq-rep-4$ values from the baseline modes regardless of embedding methods whereas our proposed model has the closest $seq-rep$ attribute to the ground truth. To further investigate the degeneration phenomenon, the step-wise probabilities assigned to the sampled tokens generated by different models and humans (ground truth) averaged for the same 10 melodic seeds are plotted in Figure~\ref{fig:probility_sampling}. It can be observed that the baseline models assign unnaturally high probabilities with low variance for sampled tokens across all time steps whereas the proposed RIPO transformer assigns probabilities similar to the ground truth (human). Hence, we may conclude that the proposed RIPO transformer can effectively tackle the degeneration problem. In order to make more fair comparisons of the objective and subject evaluation of the generated music, we use top-k sampling with sampling temperature = 1.2 for the baseline models such that the baseline models will have similar $seq-rep$ attributes to the ground truth while the sampling temperature for our proposed RIPO transformer remains to be 1.0.

\subsection{Evaluation of the generated music quality}
To reflect the music quality in objective tests, we use the following metrics: 1. the KL-divergence of the pitch ($KL_p$) and the duration ($KL_d$) generated by different models compared to the ground truth. We apply Kernel Density Estimation using Gaussian kernels to the pitch and duration histograms in order to estimate the probability distribution; 2. the in-scale ratio ($ISR$) of the generated pitches. Since the dataset is transposed to C major and A minor before training, $ISR$ is calculated as the ratio of pitches in the C major scale; 3. the arpeggio ratio ($AR$) is calculated as follows. We first split the generated music into 4-grams. In the resulting 4-grams, we obtain the number of the 4-grams with similar duration (max. 1 different duration token) and monotonically increasing or decreasing pitches (interval less than a perfect 4th). $AR$ is calculated as the number of these 4-grams divided by the total number of 4-grams. The resulting metrics are listed in Table~\ref{tab:comparison}. From the results, we observe that our proposed model almost consistently outperforms the baseline models except Model No.7 in terms of $ISR$. The $AR$ of Model No.7, however, is extremely low compared to other models. Since the metrics of the MT models consistently outperform those of the LT models, we choose Model No.1 (with the best objective metrics) as the best baseline model and use the results from Model No.1 in the listening test to compare with the results from Model No.9 (ours). 

The results of the listening test are shown in Table~\ref{tab:lst}. The metrics are obtained by taking the average of ratings across all participants. We observe that our proposed model outperforms the best baseline model in all 4 subjective metrics, indicating that RIPO transformer is able to generate music with better quality than existing SOTA transformers.

\begin{table}[h!]
  \caption{Listening test (RIPO transformer and best baseline model). Ratings are based on a 5-point Likert scale.}
  \label{tab:lst}
  \centering
\begin{adjustbox}{width=\linewidth,center}

  \begin{tabular}{ccccc}
    \toprule
    Model  &   \makecell[l]{ overall rating} &\makecell[l]{interestingness}& \makecell[l]{pitch correction} & \makecell[l]{ duration correction}  \\   
    \midrule
    MT+WE   &2.80 &2.94 &2.97&2.95 \\ 
    RIPO+FME &\textbf{3.57}&\textbf{3.52}&\textbf{3.43}&\textbf{3.37}  \\ 

    \bottomrule
  \end{tabular}
\end{adjustbox}
\end{table}

\section{Conclusion}

We propose a musically-meaningful embedding function: Fundamental Music Embedding (FME), which is based on a bias-adjusted sinusoidal encoding, and within which both the absolute and relative embedding co-exist and can be represented explicitly. Moreover, we propose a novel attention mechanism for symbolic music modeling that utilizes relative index, pitch and onset embeddings (RIPO attention). The results from our experiments show that our proposed RIPO transformer equipped with FME outperforms the SOTA transformers (i.e., music and linear transformer) using different embedding methods in a melody completion task. In a music generation task, we notice that our proposed model is the only model that is not affected by the notorious degeneration phenomenon. The music generated by our proposed RIPO transformer outperforms the SOTA transformers in both subjective and objective evaluations.



\bibliography{fme_ripo}
\section{Appendix}
\subsection{Proof of the Translational Invariance Property}

The L2 distance between $f_a$ and $f_b$ in the FME space is calculated as follows:
\begin{equation} \label{eq:a}
 \|FME_F(f_a)-FME_F(f_b)\|_2=(\sum_{k=0}^{\frac{d}{2}-1}(P_k(f_a)-P_k(f_b))^2)^\frac{1}{2}
\end{equation}
\begin{equation}\label{eq:b}
\begin{split}
(P_k(f_a)-P_k(f_b))^2 &= [sin(w_kf_a)-sin(w_kf_b)]^2\\&+[cos(w_kf_a)-cos(w_kf_b)]^2 \\
&=2-2cos(w_k(f_a-f_b))\\
&=2-2cos(w_k|f_a-f_b|) 
\end{split}
\end{equation}
In Equation~\ref{eq:b}, the trainable bias vector in $P_k(f)$ is cancelled out and $f_a-f_b$ is converted to its absolute value $|f_a-f_b|$ since the cosine function is even. By plugging Equation~\ref{eq:b} in Equation~\ref{eq:a}, we obtain:
\begin{equation} \label{eq:d}
 \|FME_F(f_a)-FME_F(f_b)\|_2=[d-2\sum_{k=0}^{\frac{d}{2}-1}cos(w_k|f_a-f_b|)]^\frac{1}{2}
\end{equation}
As a result, in Equation~\ref{eq:d} we show that for any token pair with the same relative attribute (i.e., interval = $|f_a-f_b|$), the L2 distance between the embedded tokens in the FME space is equal because the L2 distance only depends on $|f_a-f_b|$.

\begin{figure}
    \centering
    \includegraphics[width = 0.75\linewidth]{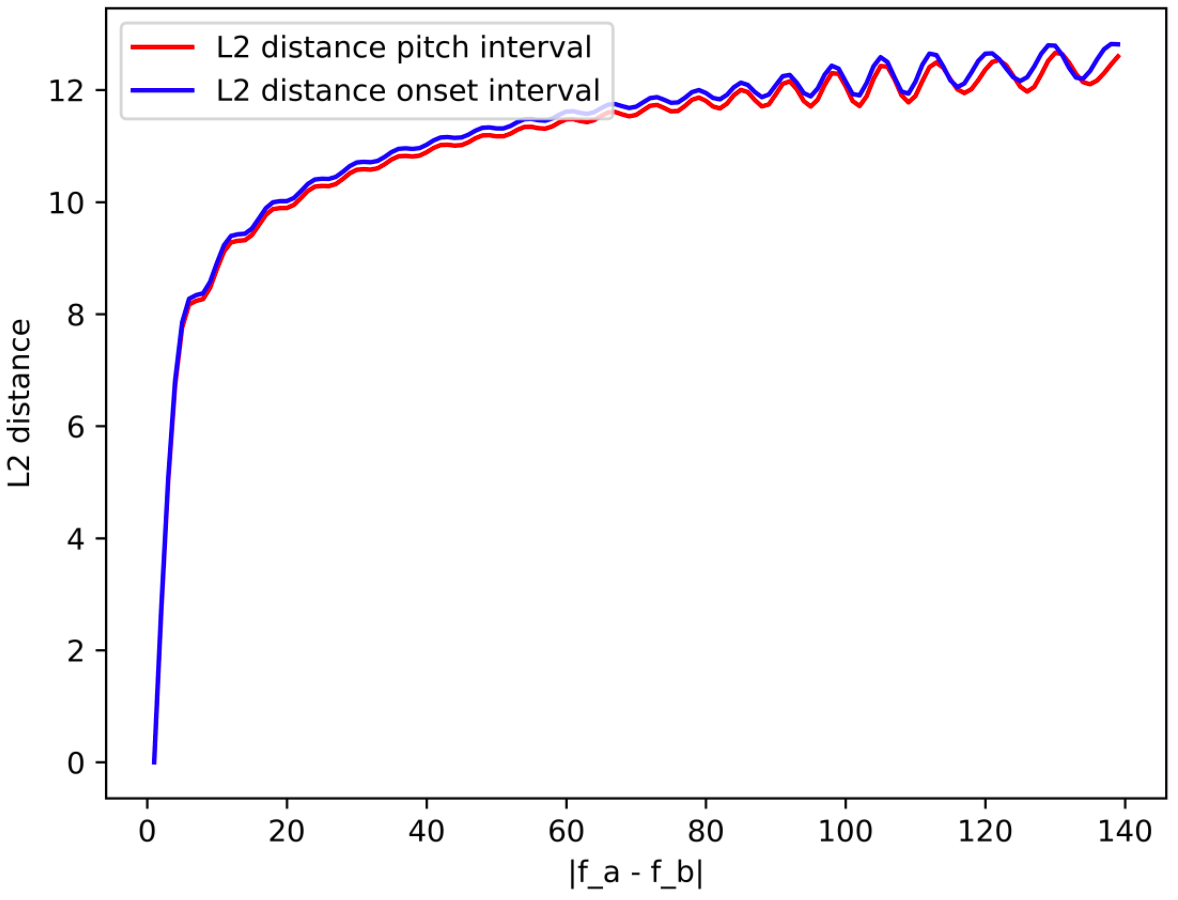}
    \caption{The relationship between $|f_a-f_b|$ and their L2 distance in the FME space.}
    \label{fig:l2dist}
\end{figure}

Next, we will discuss the relationship between $|f_a-f_b|$ and their L2 distance in the FME space. This can be interpreted as: if the interval between two pitches increases, how will the L2 distance between the pitch pair in the FME space change? Since the cosine function is 2$\pi$-periodic, in Equation~\ref{eq:d} by replacing $f_b$ with $f_b+\frac{2\pi m}{w_k}$ where $m$ is an integer, the same L2 distance would be obtained. Empirically in our experiments, we set the $B$ value in Equation 1 in Section 3 to be 9,919 and 7,920 for $FME_P$ and $FME_D$ respectively and we set $d=256$. We can hence obtain a function plot of $|f_a-f_b|$ versus $\|FME_F(f_a)-FME_F(f_b)\|_2$ in Figure~\ref{fig:l2dist}. From the figure, the L2 distance, overall, increases with an increasing interval $|f_a-f_b|$. Yet, we observe that the oscillating behaviour for large interval values.


\subsection{Proof of the Transposability Property}
In this section, we prove that in the FME space, a relative embedding $FMS_F(\Delta f)$ can be utilized to transpose any embedded FMT: $FME_F(f)$ to $FME_F(f+\Delta f)$. We first construct a transformation sub-matrix $T_k(\Delta f)$ as follows using $A_k(\Delta f)$ which is defined in Equation 4 in Section 3:
\begin{equation}\label{eq:rot}
\begin{split}
    T_k(\Delta f) &= \begin{bmatrix}
        \begin{bmatrix}0&1\\-1&0\end{bmatrix}\cdot A_k(\Delta f)^T & A_k(\Delta f)^T
     \end{bmatrix}\\
     &=\begin{bmatrix}cos(w_k\Delta f)&sin(w_k\Delta f)\\-sin(w_k\Delta f)&cos(w_k\Delta f)\end{bmatrix}
\end{split}
\end{equation}
The transformation matrix $T(\Delta f)$ is obtained by arranging the sub-matrices into a block diagonal matrix (using the \verb|torch.block_diag()| function):
\begin{equation}\label{eq:x}
T(\Delta f) =\begin{pmatrix}
 T_0(\Delta f)\\
& \ddots\\
&&T_{\frac{d}{2}-1}(\Delta f)
\end{pmatrix}
\end{equation}
We prove the transposability property in Equation~\ref{eq:y} where the bias vector $B = [b_{sin}^0, b_{cos}^0, ..., b_{sin}^{\frac{d}{2}-1}, b_{cos}^{\frac{d}{2}-1}]$ is defined in $FME_F(f)$ in Section 3 Equation 2. From Equation~\ref{eq:y}, we observe that $FME_F(f+\Delta f)$ can be directly obtained from $FME_F(f)$ and $T(\Delta f)$ (which is constructed from the $FMS_F(\Delta f)$) via a linear operation thus fulfilling the transposability property. 
\begin{equation} \label{eq:y}
    FME_F(f+\Delta f) = [T(\Delta f)  \cdot (FME_F(f) - B)^T]^T + B 
\end{equation}
Readers are invited to use the code and google colab tutorial provided in Github repository to test and explore the translational invariance and transposability property of the proposed FME.
\begin{figure*}[h!]
    \centering
    \includegraphics[width = 0.8\linewidth]{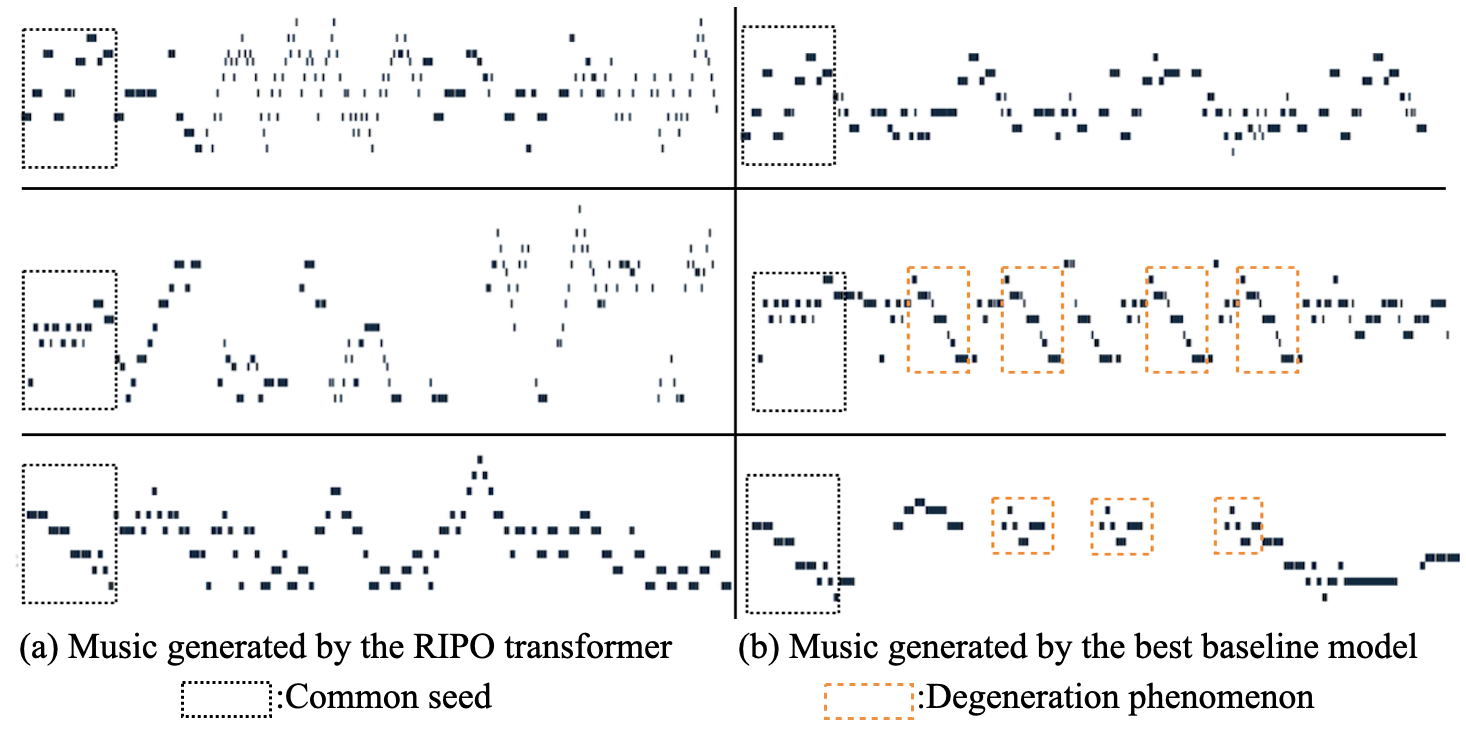}
    \caption{Music generated by the RIPO transformer and the best baseline model.}
    \label{fig:difference}
\end{figure*}

\subsection{Implementation details}
\subsubsection{Data preprocessing}
The pitch dictionary size is 131 which includes 128 MIDI pitches and 3 non-FMTs: the ``pad" token, the ``rest" token and the ``sustain" token. The duration dictionary size is 17 which includes 16 duration tokens (from a 16th note to a whole note) and the ``pad" token. 

\subsubsection{Fundamental Music Embedding and parameters}
The $B$ value in Equation 1 for $FME_P$ and $FMS_P$ is 9,919. The $B$ value for $FME_D$, $FMS_D$, $FME_O$ and $FMS_O$ is 7,920. They are set to be co-prime to each other. 

 The B values of $PE_O$ in Equation 6, $FME_O$ in Equation 7 and $FME_D$ in Equation 10 are set to be equal in order for RIPO attention layers to link the onset information (of the data) with the position information (of the model).

\subsubsection{Transformer details}
We implement our RIPO transformer using PyTorch, and we provide the code of the proposed RIPO attention in the GitHub repository. The training batch size is 16. We use Adam Optimizer with a learning rate of 0.001 with a scheduling decay. We use 2 attention layers (equipped with 8 attention heads) for all the models in the paper. The hidden dimension size for the transformer is 256.

\subsection{Visualization of the generated music}

Figure~\ref{fig:difference} shows three pairs of generated music from our proposed RIPO transformer and the best baseline model in pianoroll notation. Each pair of music shares the same melodic seed. We notice that the music generated by our proposed RIPO transformer is able to generate ``arpeggios" and interesting melodic movements even in an unconditional setting whereas the baseline models hardly generate these melodic movements. Instead, even with increased sampling temperature, the degeneration phenomenon still exists in the results generated by the baseline model. Readers are invited to listen to the samples provided in the GitHub Repository.

\subsection{Discussion}
\subsection{Relationship between FME and REMI/CP}
REMI and CP are data representations (high level) while FME is a type of embedding (low level). In other words, FME can be integrated into the existing event-based data representations by replacing WE with FME when embedding FMTs. For instance, given a REMI sequence = [... , vel(30), pitch(60), dur(4), ...]. Compared to the embedded REMI sequence: [... , WE(30), WE(60), WE(4), ...], the FME-enhanced REMI will be: [..., WE(30), FME\_p(60), FME\_d(4), ...]. 

Our data representation which is illustrated in Fig. 2, data encoding, is similar to the CP representation but without the performance attribute.


\end{document}